# Protein token: a dynamic unit in protein interactions

Si-Wei Luo[1†], Yi-Hua Jiang[1†], Zhi Liang[1*] & Jia-Rui Wu[1,2,3]

[1]School of Life Sciences, University of Science & Technology of China, Hefei, China

[2]Key Laboratory of Systems Biology, Institute of Biochemistry and Cell Biology, Shanghai Institutes for Biological Sciences, Chinese Academy of Sciences, Shanghai, China

[3]School of Life Science and Technology, ShanghaiTech University, Shanghai, China

†These two authors shared similar contributions to this study.

*Correspondence to: Z.L., E-mail: liangzhi@ustc.edu.cn

## Abstract

In this study, we introduced a new unit, named "protein token", as a dynamic protein structural unit for protein-protein interactions. Unlike the conventional structural units, protein token is not based on the sequential or spatial arrangement of residues, but comprises remote residues involved in cooperative conformational changes during protein interactions. Application of protein token on Ras GTPases revealed various tokens present in the superfamily. Distinct token combinations were found in H-Ras interacting with its various regulators and effectors, directing to a possible clue for the multiplexer property of Ras superfamily. Thus, this protein token theory may provide a new approach to study protein-protein interactions in broad applications.

## Keywords

Residue interaction union network; Small GTPase; Protein token; Protein-protein interaction

## Introduction

Proteins are macromolecules with a complicated organization, with different modules operating at different scales. Structural domains include a linear sequence of residues and operate at an independent functional scale with clear boundary in 3D structure [1]. Protein sectors comprise co-evolved amino acid sites and operate at a scattered scale [2]. Both domains and sectors represent only static arrangements of amino acid residues. Due to the dynamic nature of molecular interactions, it is hard to fully understand the conformational changes and functional state transitions of a protein involved in the interaction process using static representations alone, especially to some hub proteins [3,4]. With multiplexer property, these hub proteins usually have regulators and effectors that outnumber the substrate, while there are not as many units to handle the binding variety [3].

In this study, we address this problem by introducing a new protein unit that defines residues cooperating in the transition of protein-protein interaction. Taking the Ras super family of small GTPases as a model system, each GTPase has different functional states, including active/inactive states and states interacting with various partners. We constructed a residue interaction network (RIN) for each functional state of a protein. We then merged the RINs of two different functional states into a union network termed residue interaction union network (RIUN). By applying network analysis to the RIUN, we identified a group of spatially relevant residues cooperating in the protein state transitions, named as "protein tokens". Comparative and evolutionary analysis of different GTPases showed that protein tokens are both topologically and evolutionarily conserved. Furthermore, we investigated the token combinations of H-Ras in different states when it interacts with different partners. We found that when H-Ras interacts with each partner, the corresponding protein token is rearranged to different patterns, that distinct each other, which together form a so-called "token combination" signature of H-RAS. These protein-protein interaction tokens and the token combination are characteristic of H-RAS, and also provide an informative description for the multiplexer property of hub proteins. We propose that this concept of protein token may serve as a universal approach to investigate general protein interactions in dynamic signaling.

**Materials and Methods**

*RIUN generating*. The protein structures were downloaded from Protein Data Bank (Supplementary Information Table S1). To guarantee the reliability and consistency of subsequent analysis, in this study we only considered the small GTPases of *Homo sapiens* with structures solved by X-ray diffraction. Also, for each small GTPase investigated, the structures of different functional states were chosen if they share similar conditions. The coordinates of hydrogen atoms were added to a raw PDB file using the Reduce program [5]. The Probe program [5] was then applied to the new PDB file to determine its atomic packing. The radius of the probe was set to 0.25 Å. According to the calculated scores of noncovalent interactions, a RIN was generated for each PDB using RINerator [6]. In this weighted and undirected network, each vertex represents a residue, and each edge is assigned a weight of Probe score to represent a noncovalent interaction between two residues. The jCE algorithm [7] of RCSB Protein Comparison Tool website [8] was used to generate the structure alignment file of two target proteins. Given two RINs of the same protein in two different functional states and the structure alignment file of corresponding 3D protein structures, the RINalyzer plugin [6] in Cytoscape [9] was used to read and merge the two RINs into a union network called RIUN. In a RIUN, each vertex is a residue of the protein and each edge corresponds to a noncovalent interaction between two residues with some edges belonging to only one of the two RINs, and others belong to both.

*Molecular Complex Detection (MCODE) algorithm*. MCODE [10] is an algorithm for detecting clusters in a network, which is based on vertex weighting by local neighborhood density and outward traversal from a locally dense seed to isolate the dense regions. We applied the MCODE plugin in Cytoscape to a RIUN and obtained a set of clusters or subnetworks. Each subnetwork corresponds to a residue set and is defined as a token. In this study, we used default scoring and finding parameters that have been optimized to produce the best results for the average network [10]: degree cutoff = 2, haircut = ON, fluff = OFF, node score cutoff = 0.2 and k-core = 2.

Protein evolution conservation estimation. To calculate the evolutionary rate for each residue of H-Ras, we downloaded the sequences of human H-Ras and all its 18 orthologs in other species (*L.chalumnae, M.acridum, M.mulatta, M.musculus, N.leucogenys, O.garnettii, O.latipes, O.niloticus, P.abelii, P.alecto, P.troglodytes, R.norvegicus, S.tridecemlineatus, T.chinensis, T.guttata, T.nigroviridis, X.maculatus, X.tropicalis*) from InParanoid 8 [11]. Clustal Omega [12] was used to generate multiple sequence alignment (MSA) of the orthologous protein sequences. The evolutionary rates of H-Ras token residues and non-token residues were then estimated by using Rate4Site [13], which takes the MSA as input.

*Editing distance calculation of token combinations.* The set of tokens extracted from a RIUN form a specific token combination. The tokens can be divided into three classes according to the presence of residues belonging to protein domains: (1) Class 1 tokens are composed of both domain residues and residues not belonging to any domain, (2) Class 2 tokens are formed by more than one domain residues, (3) Class 3 tokens contain a single domain residue only. Then, we introduced edit distance to evaluate the dissimilarity between two token combinations of the same protein. Given two token combinations $C_1=\{T_1^{(1)}, T_2^{(1)},…, T_N^{(1)}\}$ and $C_2=\{T_1^{(2)}, T_2^{(2)},…, T_M^{(2)}\}$, where $T_i=\{A_{i1},…, A_{ij},…, A_{iL}\}$ represents a token and $A_{ij}$ denotes the ID number of a member residue j, we defined the following operations:

Residue deletion: remove a residue from a token.

Residue insertion: introduce a new residue to a token.

Residue substitution: change the ID number of a residue in a token.

Token deletion: remove a token from a token combination.

Token insertion: introduce a new token to a token combination.

Because it is hard to measure the free energy of an operation in physics, different operations in this study are equally important and reversible. Like the Hamming distance [14] of two strings in information theory, the editing distance between two token combinations is defined as the minimum number of operations to transform a token combination into the other, which brings the only value to measure the similarity between two different token combinations.

*Protein dynamic simulation (MD) and residue root-mean-square deviation (RMSD) calculation.* VMD [15] was used to remove water molecules from a PDB file and to generate a PSF file by using the topology file top_all27_prot_lipid.inp of the CHARMM force field [16] Then the protein structure was put into a water sphere and generated a new PDB file and a PSF file. A molecular dynamic simulation using NAMD [15] were performed by taking these files as input. During simulation, the temperature and time step were set at 310K and the 2fs/step, respectively. The average RMSD of each residue in a protein was calculated using VMD.

**Results**

Small GTPases work like molecular switches of the signal transduction systems. They are active when bound to GTP and inactive when bound to GDP [17]. Here, we investigated the conformational changes between active and inactive states for seven different Ras superfamily proteins [18], including H-Ras and Rap-2a (Ras subfamily), Rab-4a and Rab-11a (Rab subfamily), RhoA and RhoC (Rho subfamily) and Arf6 (Arf subfamily). As the workflow shown in Figure 1A, given two functional states of a protein, we first constructed a RIN for either state. Here, we generated two RINs for each small GTPase (P) studied, one for the active state (P/GTP) and the other for the inactive state (P/GDP). We then merged the two RINs into a RIUN (P/GDP, P/GTP), and extracted the topological clusters from the RIUN by using a network analysis tool. Figure 1B shows the RIUN of H-Ras. The extracted clusters composed of G domain residues were also present in figure 1C. These clusters can be divided into three classes regarding the participation of G domain residues. Class 1 clusters are hybrids of residues belonging to G domains and other parts, such as the cluster that includes the G5 domain residue A146 and other four non-G domain residues. Class 2 clusters have more than one G domain residues, such as the cluster composed of switch II residues Q61, E62 and E63. Class 3 clusters contain one single G domain residue, for example, the cluster formed by switch I residue I36 alone. Furthermore, some of these clusters have special topological structures including triangles (such as cluster of G10, A11 and K16) and hexagons (such as cluster of Y64, S65, A66, M67, R68 and D69).

Switch II domain is the main part of small GTPases and it moves upon activation by GTP [17]. And residue interaction rearrangements of its RIUN clusters exhibit more dynamic than others (Figure 1C). We examined the topological clusters covering switch II residues of different Ras superfamily proteins (Figure 2A). We found that some topological clusters, such as hexagons and triangles, are present in different Ras superfamily proteins, and proteins from the same subfamily have more similar cluster combinations. For instance, in Rab subfamily proteins, the switch-II clusters consist of one hexagonal and one triangular class 2 clusters, one or three class 3 clusters and two class 1 clusters. The switch-II clusters of Ras subfamily proteins have one hexagonal class 2 cluster in common. For Rho subfamily proteins, there is one triangular class 2 cluster and one or two class 1 clusters in the switch-II clusters. In Arf6, one hexagonal class 2 cluster is also present in two additional class 1 clusters. The proximity of Ras subfamily to Rab and Rho subfamilies in the sense of their switch-II cluster combinations agrees with their phylogenetic relationship reconstructed from MSA analysis (Figure 2A), suggesting these topological clusters are conserved in protein evolution. Since the edges correspond to residue interactions, the topological clusters capture the cooperation between residues involved in conformational changes of small GTPase state transition (Figure 2B). Hereafter, we termed these conserved sub-domain topological structures "protein tokens".

To get a further understanding of protein tokens, we investigated tokens covering different H-Ras domains involved in transitions between the inactive state and active states when complexing with Byr2 and PI3K [19], respectively. The G1 domain tokens of (H-Ras/GDP, H-Ras/GTP) and (H-Ras/GDP, H-Ras/Byr2) have two triangles, while those of (H-Ras/GDP, H-Ras/PI3K) include only one triangle (Figure 2C). The switch II tokens of (H-Ras/GDP, H-Ras/GTP) include one triangle and one hexagon, but those of (H-Ras, H-Ras/Byr2) and (H-Ras, H-Ras/PI3K) contain one pentagon and one triangle or quadrilateral (Figure 2D). These results reveal that these conserved topological structures exist in different domains of a protein and rearrange with different interacting partners. We analyzed the motion of residues between the inactive and active states by calculating $|\Delta RMSD|$, the absolute value of residue-wise RMSD difference, in molecular dynamics simulation. It was found that two peaks are present on the G1 domain $|\Delta RMSD|$ profile of (H-Ras/GDP, H-Ras/Byr2), while there is only one peak for (H-Ras/GDP, H-Ras/PI3K) (Figure 2C). And all the switch II $|\Delta RMSD|$ profiles have two peak regions (Figure 2D). The number of peak regions matches the number of tokens, and the token patterns change with the $|\Delta RMSD|$ profiles of a domain. This observation means that protein tokens capture conformational changes associated with protein interactions. We also calculated the evolutionary rates of H-Ras token residues and non-token residues (Supplementary Information Figure S1). We found the average evolutionary rate of token residues is slower than those of non-token residues (p= 0.00038) and all residues (p=0.0027) of H-Ras, which indicates the conservation of token residues. All the above results indicate that protein tokens are units for protein conformational changes in state transition involved in protein interactions with topological and evolutionary conservation.

As a hub protein in the protein interactome, H-Ras works in the manner of a multiplexer. That is a particular GEF activates the small GTPase, which in turn selects and binds a particular effector [3]. We analyzed several effectors, including Byr2, NORE1A, RalGDS and PLC-ε, which interact with H-Ras through their Ras binding domains (RBDs) [19]. The secondary structure alignment of these RBDs exhibits a high degree of similarity (Figure 3A). This observation suggests that the multiplexer property of H-Ras requires more than static structure recognition, which means the multiplexer property of H-Ras could not be simply explained in terms of functional domains. Considering that small GTPases rely on the intrinsic plasticity of their structures combined with stereochemical properties of residues [3], we resorted to the concept of protein tokens. We generated the protein tokens of (H-Ras/GDP, H-Ras/Partner), where Partner includes the RBDs of effector Byr2, NORE1A, RalGDS, PLC-ε, and PI3K-γ as well as a regulator SOS-1 [19] (Supplementary Information Figure S2~S7). Tokens of H-Ras functional domains were divided into three classes of residue sets (Figure 3B). We calculated the edit distances between these different token combinations and found that the token combination of H-Ras/Byr2 RBD complex resembles that of H-Ras/SOS-1 complex than other complexes (Figure 3C). In cellular signal transduction, the main effector of SOS-1-activated H-Ras is Raf [19] and Byr2 is a Raf homologous protein. Therefore, this result reveals a possible connection between protein tokens and PPI preference that distances between different token combinations may represent the multi-specificity of a hub protein as a multiplexer in PPIs.

**Discussion**

According to the above results, protein tokens are conserved units associated with protein conformational changes involved in interactions, and their combinations provide an informative representation of interaction specificity. Unlike backbone-based sequential representation of domains or sectors, tokens come from residue cooperation in protein non-covalent interaction networks, which means protein tokens may be more suitable for describing protein state changes in interactions. Assuming that different residues compose an alphabet and protein tokens are code words over this alphabet, a functional state of a protein may serve as an encoding system by the combination of different protein tokens. Based on this assumption, we may further access PPIs in signal transduction as a maximum likelihood decoding process of the hub protein just like the coding theory of artificial communication system [20].

By analyzing contacts or interactions between residues, some previous works had investigated the conformation changes of H-Ras [21,22]. Despite of the difference in constructing residue interaction networks, they share some similar observations with our research. For example, residue interactions occurred more frequently with highly coevolved residues, such as V9, Y32, I36, A59, Q61, E63 and Y71 [21], which are also part of protein tokens.

The limitation of protein tokens is that they are hard to be manipulated to modulate the functions of a protein in experiments. Because protein tokens are based on residue cooperation in a conformational change, it cannot be simply removed or substituted like domains or sectors. Artificially change of protein tokens means to redesign RIN of a protein, and will needs further research in future.

Furthermore, according to the latest version of COSMIC database [23], 29 residues of H-Ras have cancer-related mutations. We found that 14 of them are covered by protein tokens. This result indicates that protein tokens are of functional significance and may bring a new approach to study the molecular mechanisms of diseases in a broad concept.

## Acknowledgements

This work was supported by grants to JRW from grants of the National Natural Science Foundation of China (31130034 and 31470808), and from a grant of Strategic Priority Research Program of the Chinese Academy of Sciences (XDA12000000); by grants to ZL from the Ministry of Science and Technology (2012CB917200, 2014CB910600).

## Author contributions

In this study, Siwei Luo conceived the idea of "protein token". Siwei Luo designed the research. Siwei Luo and Yihua Jiang carried out the computing and data analyzing. Zhi Liang and Jiarui Wu supervised the project. Siwei Luo, Yihua Jiang and Zhi Liang wrote the manuscript.

*Conflict of Interest:* none declared.

**Supporting information**

**Table S1**. The list of PDB structure files used in this study.

**Figure S1**. Revolutionary rates of H-Ras residues.

**Figure S2~ S7**. MCODE clustering results of different (H-Ras/GDP, H-Ras/Partner) RIUNs.

**Figure legends**

**Figure 1. (A)** Workflow of RIUN processing. Given two functional states of a protein, a RIN is constructed for either state by using RINerator. The two RINs are merged into a union network RIUN, in which each vertex is a residue of the protein, and each edge corresponds to a noncovalent interaction between two residues. In the RIUN, some edges belong to only one of the two RINs (red or blue edges), and others belong to both (gray edges). MCODE is then used to extract clusters, protein tokens, from the RIUN. **(B, C)** The RIUN of H-Ras (B) and the clusters composed of G domain residues (C). Cyan circle represents the residues that do not belong to G domains. Blue, red and gray

edges represent residue interactions present in an inactive state, active state and both states, respectively.

**Figure 2. (A)** Topological clusters covering the switch II domain of Ras superfamily. The phylogenetic tree presents the evolutionary relationships between the proteins. It was generated by using the Clustal Omega program, and the protein sequences used for multiple sequence alignment were extracted from the Uniprot database. In RIUN clusters, blue diamonds, yellow triangles and cyan circles denote residues belonging to the switch II, switch I and other non-domain parts, respectively. Blue, red and gray edges represent residue interactions present in an inactive state, active state, and both states, respectively. **(B)** Conformations of switch II in inactive (blue) and active (red) states of Rab-4a (PDB entry: 2mbd, 2bme), RhoA (PDB entry: 1ftn, 1a2b), and Arf6 (PDB entry: 1e0s, 2j5x). These structures are rendered using Chimera. Protein tokens covering the **(C)** G1 and **(D)** switch II domains of H-Ras in different interaction states and their |ΔRMSD| profiles.

**Figure 3. (A)** Secondary structure alignment (S: β-sheet, H: α-helix) of RBD domains of Byr2 (PDB entry: 1i35), NORE1A (PDB entry: 3ddc), RalGDS (PDB entry: 1lxd) and PLC-ε (PDB entry: 2byf). The capital red H represents the conserved helix structure across RBDs, the small gray h represents the helix structure that is not conserved across RBDs. **(B)** Protein token combinations of different active interaction states of H-Ras. The numbers indicate different residues of H-Ras. **(C)** Similarities of different token combinations based on editing distances. The similarity between the token combinations of H-Ras/Byr2 (Raf homologous protein) RBD complex and H-Ras/SOS-1 complex may indicate the PPI preference (red arrow) in this signal transduction system.

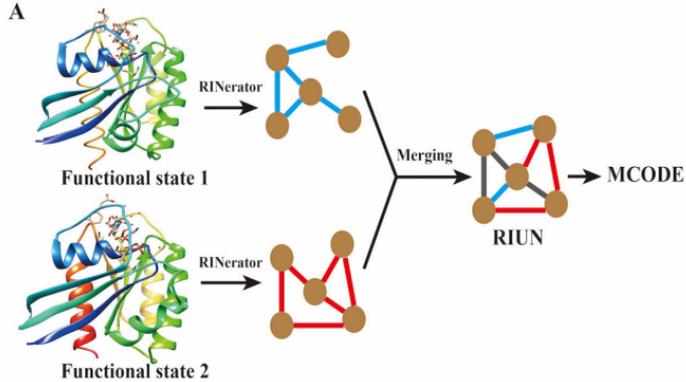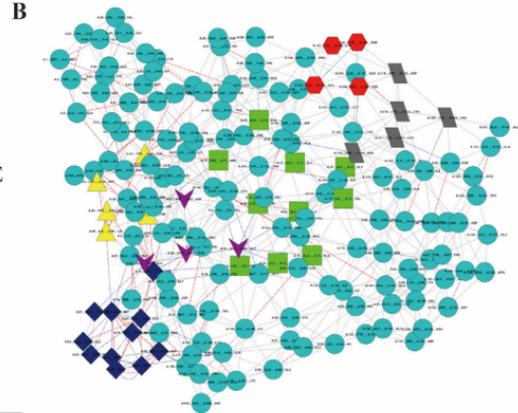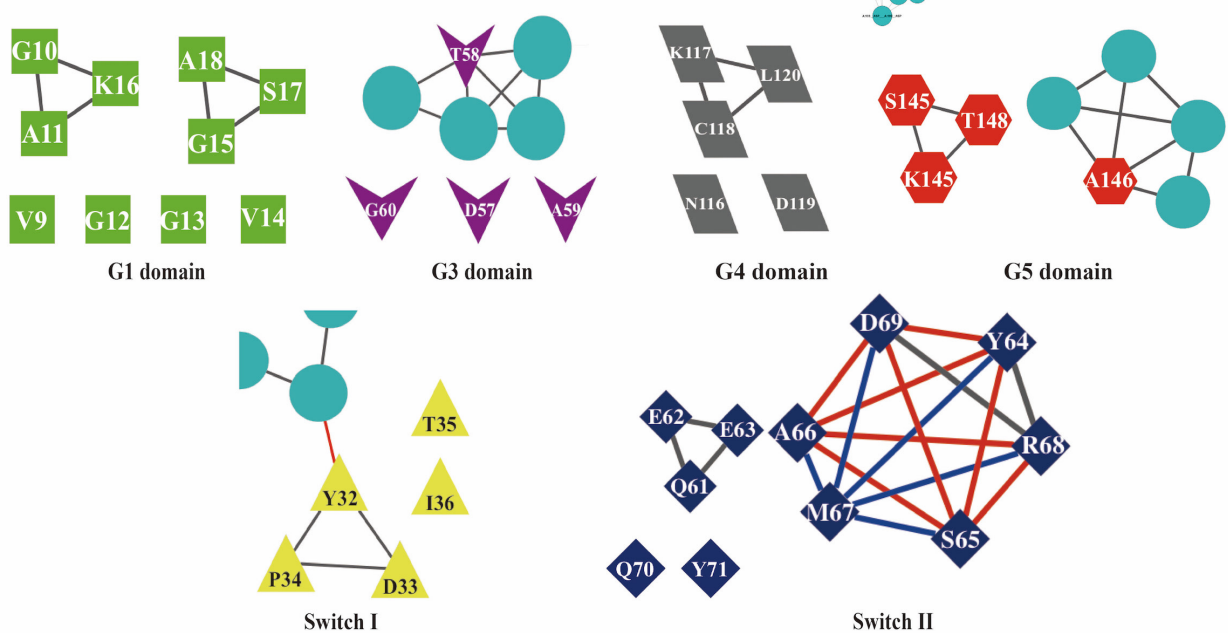

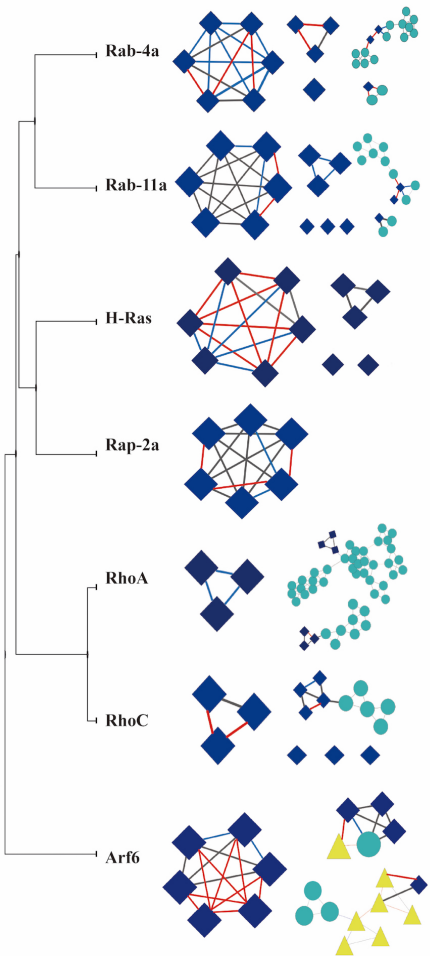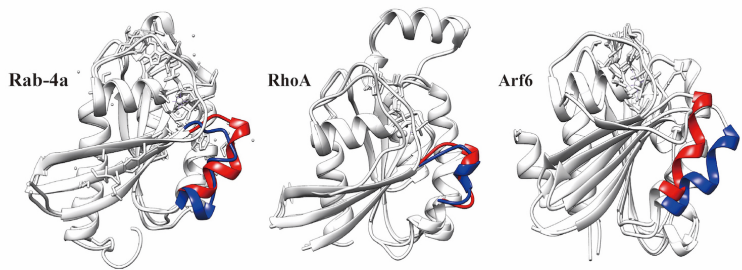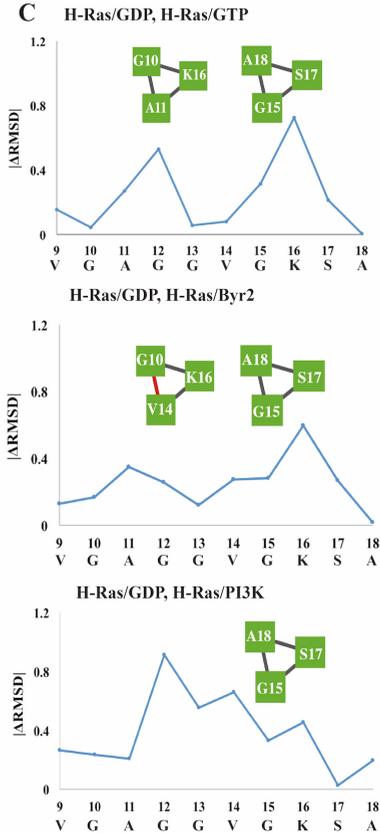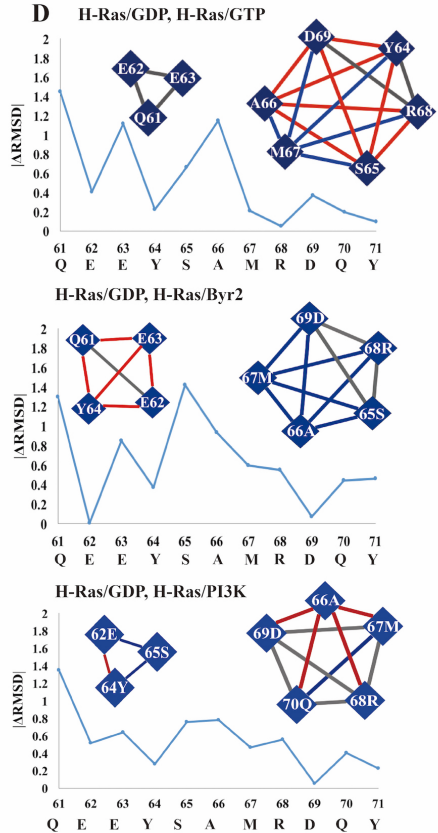

**A**

| | | | | | | | | | | |
|---|---|---|---|---|---|---|---|---|---|---|
| Byr2 RBD | - | - | S | S | H | h | S | S | H | S | h |
| NORE1A RBD | h | h | S | S | H | - | S | S | H | S | - |
| RalGDS RBD | - | - | S | S | H | - | S | S | H | S | - |
| PLC-ε RBD | - | - | S | S | H | h | S | S | H | S | - |

**B**

| | Class 1 | Class 2 | Class 3 |
|---|---|---|---|
| H-Ras/SOS-1 complex | {9,10,116,117,146} {11,14,16,35,36,57} {58,71} {145,147,148} | {15,17,18} {62,63,64} {65,66,67,68,69,70} | {12} {13} {32} {33} {34} {35} {59} {60} {61} {118} {119} {120} |
| H-Ras/PLC-ε complex | {57,58} {116,117,146} {145} | {9,10,11,16,60} {15,17,18} {62,63,64,65} {66,67,68,69,70,71} | {12} {13} {14} {32} {33} {34} {35} {36} {59} {61} {118} {119} {120} {147} {148} |
| H-Ras/Byr2 complex | {71} {58} {10,14,16} {116,117,118,120,145,146,147,148} | {15,17,18} {61,62,63,64} {65,66,67,68,69} | {9} {11} {12} {13} {32} {33} {34} {35} {36} {57} {59} {60} {70} {119} |
| H-Ras/RalGDS complex | {11} {15,17,18} {9,10,14,16} {58} {116,145,146,147,148} | {62,63,64,65,66} {67,68,69,70,71} {117,118,120} | {12} {13} {32} {33} {34} {35} {36} {57} {59} {60} {61} {119} |
| H-Ras/PI3K-γ complex | {9} {11} {35,36} {12,32,33,34,58,59,60,61,71} {116,145,146,147,148} | {15,17,18} {62,64,65} {66,67,68,69,70} {117,118,120} | {10} {13} {14} {16} {57} {60} {61} {63} {71} {119} |
| Hras/NORE1A complex | {10,14,16} {9,57,71,116} {33,35,58} {146} {118,145,147,148} | {15,17,18} {12,60,61} {62,63,64,65} {66,67,68,69,70} | {11} {13} {32} {34} {36} {59} {117} {119} |

**C**

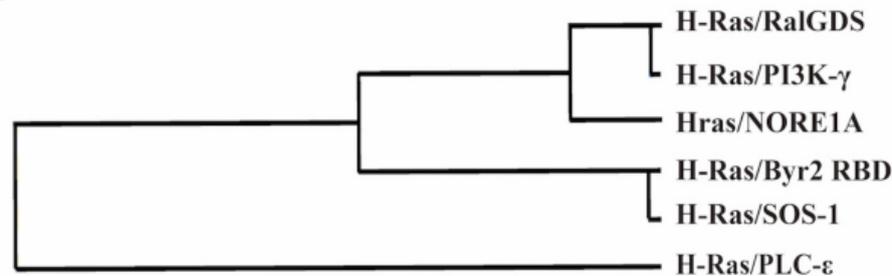

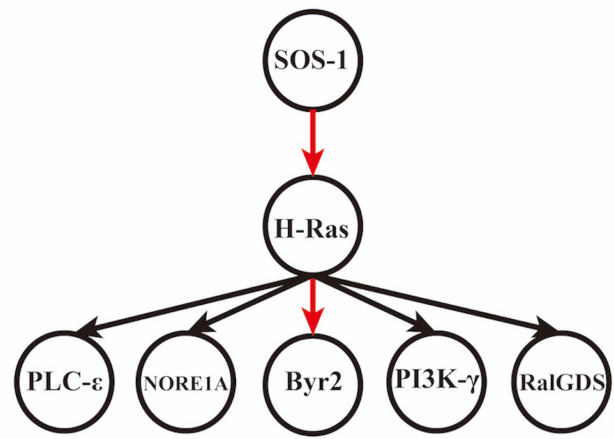

# Supporting Information

**Table S1. The PDB entries of protein structures used in this research**

| H-Ras/GDP | 4Q21 | Arf6/GDP | 1E0S |
|---|---|---|---|
| H-Ras/GTP | 5P21 | Arf6/GTP | 2J5X |
| Rap-2a/GDP | 1KAO | H-Ras/SOS-1 | 1XD2 |
| Rap-2a/GTP | 2RAP | H-Ras/ PLC-ε | 2C5L |
| Rab-4a/GDP | 2BMD | H-Ras/Byr2RBD | 1K8R |
| Rab-4a/GTP | 2BME | H-Ras/PI3K | 1HE8 |
| Rab-11a/GDP | 1OIV | H-Ras/RalGDS | 1LFD |
| Rab-11a/GTP | 1OIW | H-Ras/NORE1A | 3DDC |
| RhoA/GDP | 1FTN | Byr2 RBD | 1I35 |
| RhoA/GTP | 1A2B | RalGDS RBD | 1LXD |
| RhoC/GDP | 2GCN | PLC-ε RBD | 2BYF |
| RhoC/GTP | 2GCP | | |

**Figure S1. Evolutionary rates of H-Ras residues**

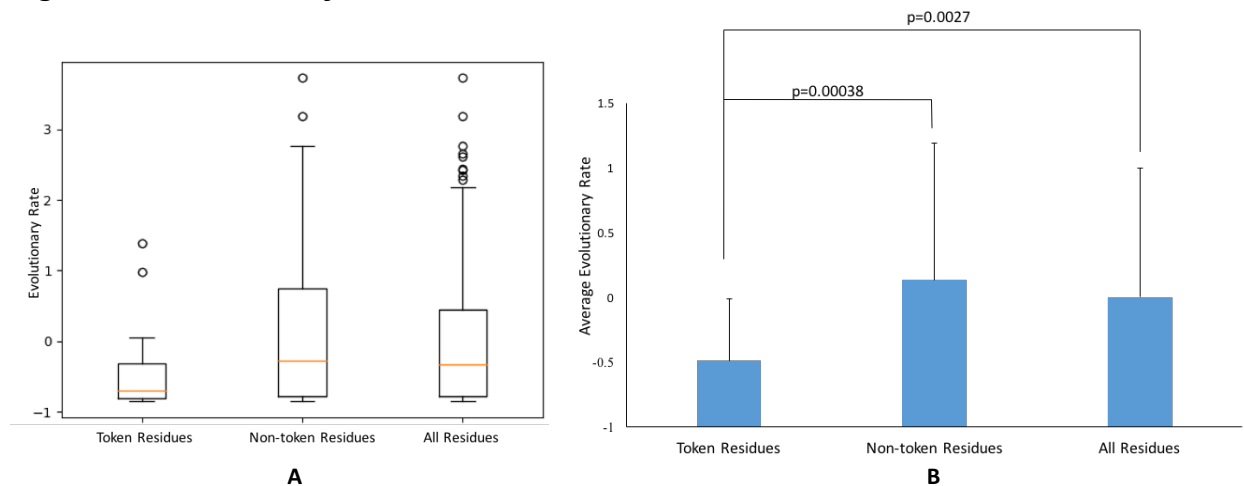

Figure S1. (A) Distribution of evolutionary rates of token residues, none-token residues and all residues of H-Ras. (B) Average evolutionary rates of token residues, non-token residues and all residues of H-Ras (p: p-value of t-test).

**Figure S2. MCODE clustering result of the (H-Ras/GDP, H-Ras/Byr2RBD) RIUN**

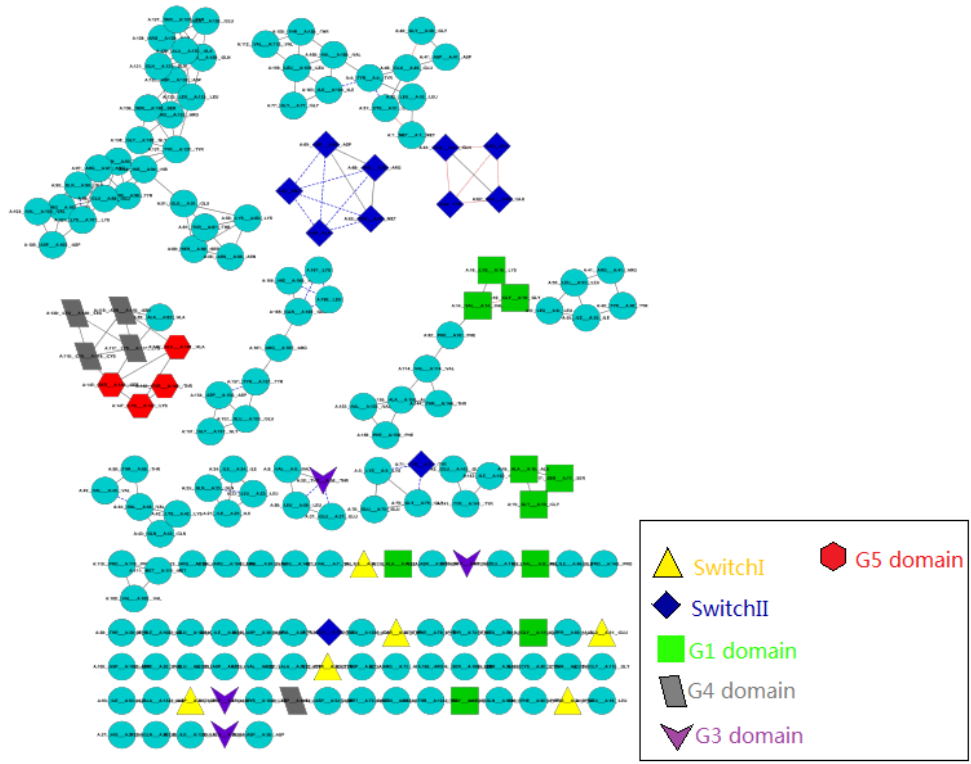

**Figure S3. MCODE clustering result of the (H-Ras/GDP, H-Ras/NORE1A) RIUN**

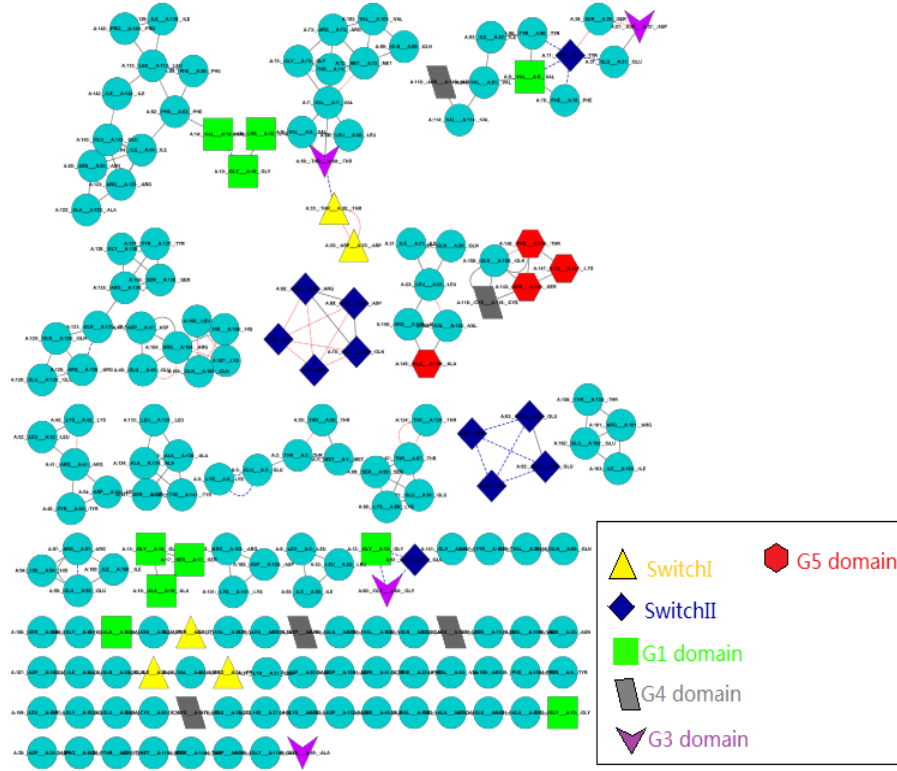

**Figure S4. MCODE clustering result of the (H-Ras/GDP, H-Ras/RalGDS) RIUN**

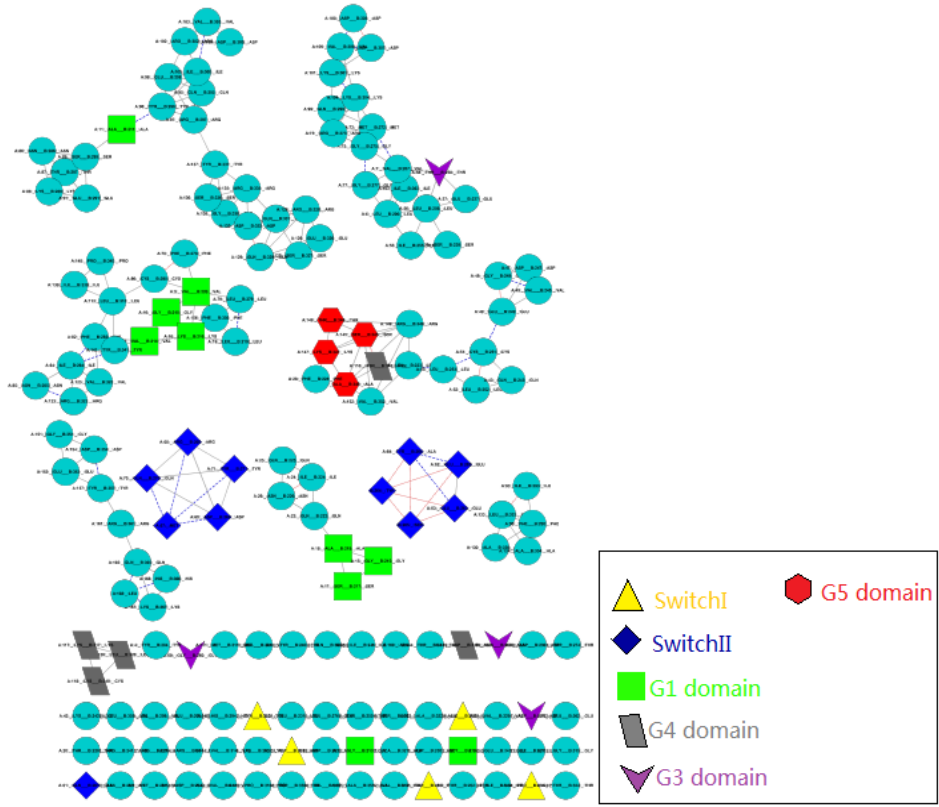

**Figure S5. MCODE clustering result of the (H-Ras/GDP, H-Ras/PLC-ε) RIUN**

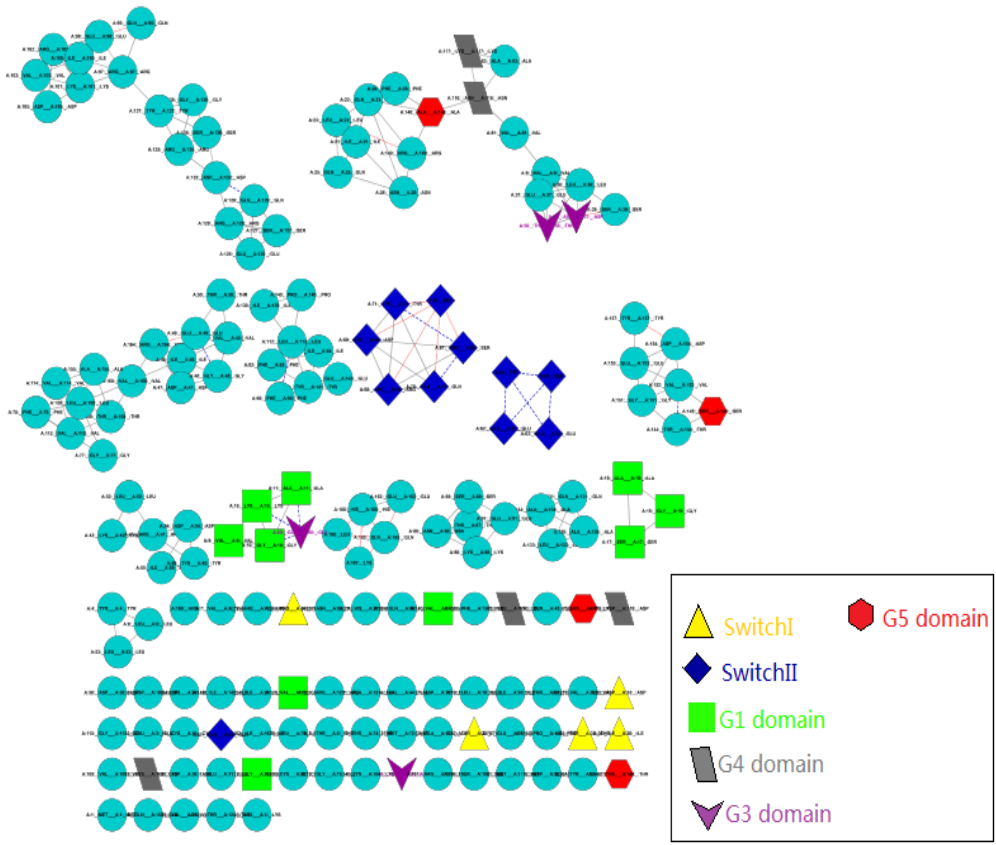

**Figure S6. MCODE clustering result of the (H-Ras/GDP, H-Ras/PI3K) RIUN**

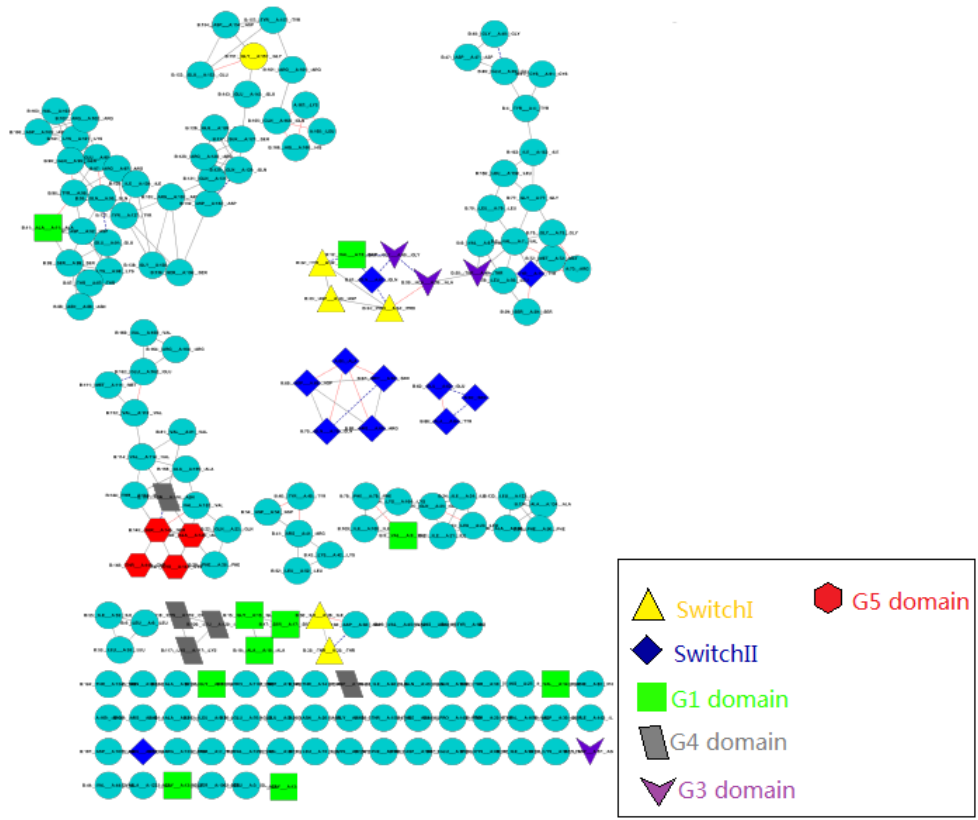

**Figure S7. MCODE clustering result of the (H-Ras/GDP, H-Ras/SOS-1) RIUN**

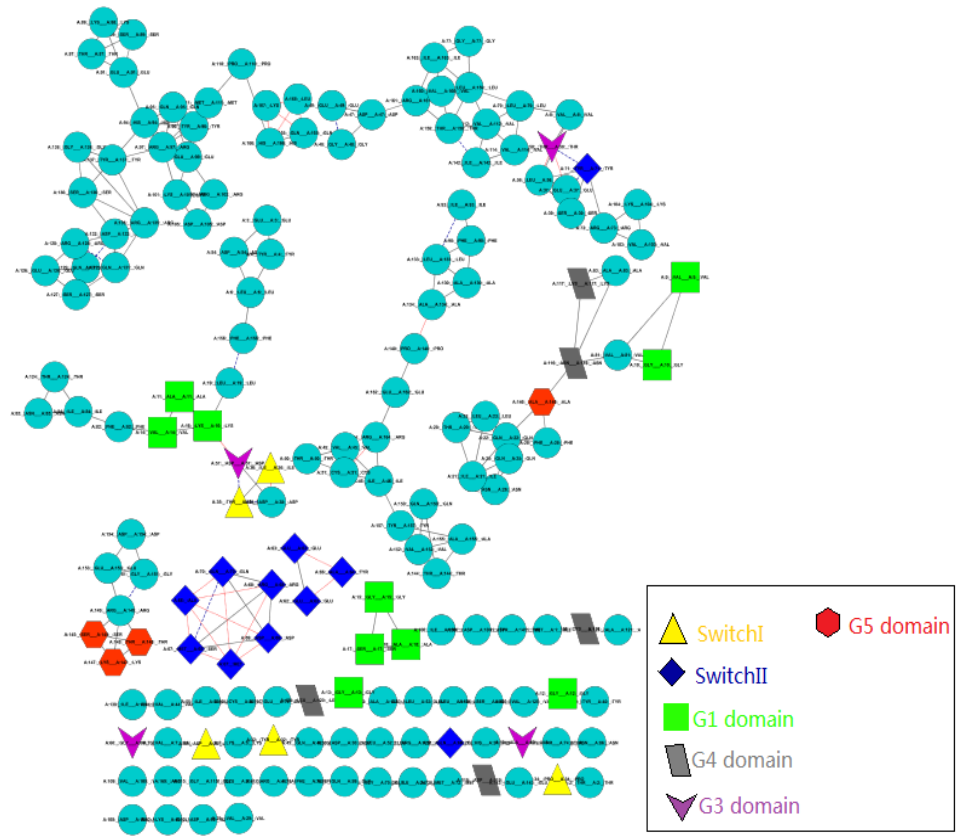